\documentclass[useAMS,usenatbib,fleqn,usedcolumn]{mnras}

\usepackage[british]{babel}             % British English hyphenation
\usepackage{graphicx}                   % Including figures

\usepackage{bm}
\usepackage{amsmath}
\usepackage{amssymb}
\usepackage{multirow}

\renewcommand{\vec}[1]{\ensuremath{\bm{{#1}}}} 
\newcommand{\mat}[1]{\bm{\mathrm{#1}}} 
\newcommand{\unit}[1]{\ensuremath{\,\mathrm{#1}}}
\newcommand{\diff}{\ensuremath{\mathrm{\,d}}}

\newcommand{\pderiv}[2]{\ensuremath{\frac{\partial #1}{\partial #2}}}

\title[3D Gravito-turbulence]{Characterizing gravito-turbulence in 3D: turbulent properties and stability against fragmentation}
\author[Booth \& Clarke]{Richard A. Booth\thanks{E-mail: rab200@ast.cam.ac.uk}$^1$ and Cathie J. Clarke$^1$ \\
$^{1}$Institute of Astronomy, University of Cambridge, Madingley Road, Cambridge, CB3 0HA, United Kingdom}

\begin{document}

\date{Accepted . Received ; in original form }

\pagerange{\pageref{firstpage}--\pageref{lastpage}} \pubyear{2015}

\maketitle

\label{firstpage}

\begin{abstract}
We have investigated the properties of gravito-turbulent discs in 3D using high-resolution shearing-box simulations. For large enough domain sizes, $L_y \gtrsim 60H$, the disc settles down into a quasi-steady state, showing no long term trends in properties or variation with box size. For smaller boxes, we find that the azimuthal wavelength of the dominant spiral modes are limited to the domain size. This is associated with a bursty behaviour that differs from the quasi-steady dynamics at larger sizes. We point out that a similar transition may be expected in global simulations at the point where the range of azimuthal wavelengths is limited by the finite disc size.  This condition (i.e. when $60 H \sim 2 \upi R$, i.e. $H/R \sim 0.1$) correctly predicts the transition to bursty behaviour previously found in global simulations for disc-to-star mass ratios in excess of 0.25. We recover a transition in the dynamics from two- to three-dimensional behaviour, characterized by a turbulence that becomes more isotropic on small scales. This turbulence likely plays an important role in the evolution of dust in self-gravitating discs, potentially dominating the collision velocity for particles with Stokes number $< 1$. Finally, we consider the stability of gravito-turbulence against fragmentation, finding that discs which cool faster than a few dynamical times fragment immediately, supporting previous results. We also find hints of stochastic fragmentation at marginally longer cooling times, in which a fragment forms from a quasi-equilibrium state. However, this makes little practical difference to region where planet formation via gravitational instability may occur. 
\end{abstract}


\section{Introduction}
\label{Sec:Intro}

Early in their evolution, protoplanetary discs pass through a phase in which their mass is comparable to that of the central protostar. During this phase, the disc's own gravity plays an important role in its evolution, with self-gravitating modes transporting angular momentum, driving accretion through the disc. The outcome of this evolution depends on the balance of heating and cooling: if the cooling is not too fast then the disc is expected to maintain marginal stability, where the Toomre's $Q \approx 1$ \citep{Safronov1960,Toomre1964,Paczynski1978}, with the heating produced by the spiral shocks being balanced by cooling (so-called gravito-turbulence). Conversely, when cooling is sufficiently fast relative to the orbital time-scale the discs may instead fragment. Both of these regimes have important implications for planet formation.

Fragmentation due to gravitational instability (GI) has been suggested as a way to produce giant planets \citep{Cameron1978,Boss1998}. The requirement of short cooling times, $\beta = t_c \Omega \lesssim 3$\footnote{The discussion of fragmentation here assumes $\gamma = 5/3$. Using $\gamma = 1.4$ instead increases the critical cooling time-scale by a factor of a few, but does not otherwise change the results.} (where $\Omega$ is the Keplerian angular speed and $t_c$ is the cooling time, \citealt{Gammie2001}), limits the process to the outer regions of protoplanetary discs \citep{Rafikov2005,Clarke2009,Rafikov2009}. Here, the fragment masses exceed several Jupiter masses, with brown dwarves or low mass-stars potentially being a more natural outcome \citep{Stamatellos2009,Rice2015}. However, high resolution simulations suggest that fragmentation may occur stochastically at longer time scales \citep{Paardekooper2011,Meru2011}, which would push the fragmentation limit to small radii and allow the formation of lower mass objects. 

While the boundary between immediate fragmentation and quasi-stability at $\beta \approx 3$ is becoming well understood \citep{Kratter2011, Young2015, Deng2017}, the role of stochastic fragmentation is harder to determine directly due to the need for high resolution and long integration times. However, insights from two-dimensional simulations suggest that it is unlikely that stochastic fragmentation can be important at very long cooling times for two reasons. Firstly, when $\beta \gtrsim 3$ bound clumps collapse on a cooling time, and thus may be disrupted when passing through shocks before they can contract \citep{Kratter2011,Young2015}. \citet{Young2016} showed that clumps typically encounter shocks once every few orbits, with the probability of avoiding shocks for a long time falling off exponentially. Secondly, at long cooling times the formation of bound clumps is increasingly rare \citep[e.g.][]{Paardekooper2011}. Here we use shearing-box simulations to argue that similar results must hold in three dimensions.

The gravito-turbulent regime at moderate to long cooling times has another role to play in planet formation,  through its effects on dust grains. Since dust grains drift towards the transient pressure maxima associated with the spiral arms, this provides a way to concentrate dust grains \citep{Rice2004}. For large enough sizes (around decimetre), this can result in the dust reaching high enough densities that the dust itself becomes self-gravitating, potentially producing planetesimals by direct collapse in the dust layer \citep{Rice2006,Gibbons2014}. Traditionally the strong large-scale motions in gravito-turbulence have been considered a problem for dust growth in such discs, for example with collisions between planetesimals being destructive \citep{Walmswell2013}. However, recent 2D simulations have shown that for grains with small enough Stokes number, $St = t_s \Omega \lesssim 3$ (where $t_s$ is the time-scale over which drag forces damp the relative velocity between the gas and dust), decouple from the large scale flow and may avoid fragmenting \citep{Booth2016,Shi2016}.

The dynamics on scales smaller than the disc scale height turns out to be crucial for understanding whether dust trapping leading to planetesimal formation can occur. For example, small-scale turbulence is important because it drives diffusion. This diffusion is not only important for determining the strength and extent of dust trapping, but will act against gravity, setting a minimum density required for the dust trapped in spiral arms to become self-gravitating. This in turn controls the effective Jean's mass of the clumps that form and thus the properties of planetesimals formed via this mechanism. 

Perhaps the clearest test of the small-scale properties of gravito-turbulent discs comes from studies of the Large Magellanic Cloud (LMC): the power spectrum of density fluctuations in the LMC shows a break at length scales around the disc's scale-height \citep{Elmegreen2001,Block2010}. Similar behaviour has also been seen in M33, supporting this idea \citep{Combes2012}. Indeed, using adaptive mesh refinement simulations of isolated disc-galaxies, \citet{Bournaud2010} reproduced the break in the power spectrum of density fluctuations and showed that it is associated with the scale on which gravito-turbulence transitions from being two-dimensional to three-dimensional. 

However, the applicability of galactic gravito-turbulence to the protostellar case is not immediately obvious. While \citet{Bournaud2010} show that the additional physics such as supernova feedback and dark matter are important to the dynamics, they also showed that these processes do not play a dominant role in driving the turbulence. Perhaps the biggest difference is the gas cooling time, which is much smaller than the orbital time in the galactic context, resulting in an effectively barotropic equation of state. As discussed above, the short effective cooling time makes the spirals highly unstable to fragmentation. % Baroclinic term?
Although, the power spectrum of density fluctuations is known to fall off on scales less than about the disc scale-height in the protostellar case too \citep[e.g.]{Boley2006,Cossins2009}, this is seen in both two- and three-dimensional simulations \citep[e.g.][]{Gammie2001,Cossins2009,Rice2016}. Thus it cannot be related to a transition between two- and three-dimensional dynamics, and instead is associated with the preferred scale of the dynamics. 

However, there are some hints that gravito-turbulence in protostellar discs has similarities to the galactic case. For example, \citet{Riols2017} showed that on large scales the power spectrum of kinetic energy is close to the $k^{-5/3}$ scaling expected for both two- and three-dimensional turbulence \citep{Kraichnan1967,Kritsuk2007}, as reported by \citet{Bournaud2010}. While \citet{Riols2017} saw a steeper decline in kinetic energy on scales below the scale-height, we will argue here that this is due to numerical dissipation. Furthermore, high resolution 2D simulations of gravito-turbulent protostellar discs show the formation of large-scale vortices, as expected for 2D \citep{Mamatsashvili2009,Gibbons2015,Booth2016}, although self-gravity does modify the dynamics, preventing the formation of arbitrarily large vortices \citep{Mamatsashvili2009}.

In this work we demonstrate that self-gravitating protostellar discs do transition from two- to three-dimensional dynamics, showing that this occurs on smaller scales than the most-unstable wavelength. Section 2 describes the numerical methods used. In section 3, we discuss the large-scale properties of the turbulence before considering the turbulent spectra. In section 4, we report the results of an investigation into fragmentation. In section 5 we discuss the implications of the turbulence for the dynamics of dust in gravito-turbulent discs. Finally in section 6 we summarize the results. 

\section{Numerical Method}

We employ the Godunov code \textsc{Athena} \citep{Stone2008} in order to simulate gravito-turbulence in 3D, following closely the methods used by \citet{Shi2014}. We solve the equations of hydrodynamics in a local frame co-rotating with the background flow using the stratified shearing-box approximation \citep{Hill1878,Hawley1995,Stone2010}. The equations solved are those of self-gravitating hydrodynamics,
\begin{align}
 \pderiv{\rho}{t} + \nabla \cdot (\rho \vec{v}) =& 0, \\
 \pderiv{(\rho \vec{v})}{t} + \nabla \cdot (\rho \vec{v}\vec{v} + P\mat{I} + \mat{T}) =& - 2\rho \Omega \vec{\hat{z}} \times \vec{v}  + 2 \rho q \Omega^2 x \vec{\hat{x}} \nonumber \\ 
	  & \quad- \rho \Omega^2 \vec{z}, \\
  \pderiv{E}{t} + \nabla \cdot \left[(E + P)\vec{v}\right] =& 
	\dot{Q}_{\rm cool} - \rho \vec{v} \cdot \nabla \Phi \nonumber \\
	& \quad + \rho \Omega^2 \vec{v} \cdot (2 q \vec{\hat{x}} - \vec{z}), \\
\nabla^2 \Phi =& 4 \upi G \rho,
\end{align}
where $\Phi$ is the gravitational potential due to disc and the total energy, $E$, includes the thermal and kinetic energy, ${E = \frac{1}{2}\rho v^2 + P / (\gamma - 1)}$, $\Omega$ is the local angular speed and 
\begin{equation}
q = - \pderiv{\log{(\Omega)}}{\log{(r)}} = 3/2,
\end{equation} where the final equality is for Keplerian motion. We parametrize the cooling rate in the standard way for simulations of self-gravitating discs \citep[e.g.][]{Gammie2001,Rice2003,Lodato2004,Mejia2005,Cossins2009,Meru2011,Paardekooper2011}, by setting the cooling time-scale to a multiple of the local dynamical time-scale, $t_c = \beta \Omega^{-1}$, and setting $\dot{Q}_{\rm cool} = - U / t_c$, where $U = P / (\gamma -1)$ is the internal energy per unit volume and $\gamma = 5/3$ is the ratio of specific heat capacities. The gravitational stress tensor, $\mat{T}$, is given by
\begin{equation}
 \mat{T} = \frac{1}{4 \upi G} \left[ \nabla \Phi \nabla \Phi - \frac{1}{2}( \nabla \Phi \cdot \nabla \Phi) \mat{I} \right].
\end{equation}

We solve the above equations using van Leer integration \citep{vanLeer2006,Stone2009} along with piece-wise linear reconstruction with a Courant number $C=0.4$. The gravitational potential, $\Phi$, is solved for using fast Fourier transforms using the routines in the public version of \textsc{Athena} \citep{Koyama2009,Kim2011}. Since in the shearing box approximation the simulation domain is only shear-periodic it is necessary to first shift the density back to the nearest time at which the simulation was strictly periodic, solve for $\Phi$ and then shift back to current time. We perform the shift in real space as in \citet{Gammie2001}, rather than shifting in Fourier-space \citep{Johansen2007}. 

We use the standard boundary conditions for shearing boxes in $x$ and $y$ --- periodic in $y$ and shear-periodic in $x$ --- along with outflow boundary conditions in $z$. Since our boxes are stratified, we found greatest accuracy when extrapolating the density using the hydrostatic equilibrium profile. We did this assuming the temperature in the boundary cells is the same as the last active cell. Additionally, we copied the velocities from the final cell, but set the velocity normal to the grid to zero if it was directed into the simulation box. 

As reported by \citet{Shi2014}, we found that it was necessary to introduce a density floor into the simulations as the outflow boundary conditions occasionally produce very low densities in a few cells near the upper and lower boundaries, which ended up becoming very hot and forcing the use of tiny time-steps. We typically set the threshold to $10^{-4}$ of the mid-plane density as in \citet{Shi2014}, \citet{Riols2017} and \citet{Baehr2017}. Due to mass loss through the boundary, at each time step we typically rescale the density by a constant factor to renormalise the density to its initial value, while keeping the per-unit mass quantities fixed (i.e. energy, velocity). The mass and energy added through this process is approximately a few per cent per cooling time.

For the simulation units, we scale times to $\Omega^{-1}$ and lengths to
\begin{equation}
H = \frac{\upi G \Sigma}{\Omega^2},
\end{equation}
which can be interpreted as the tidal radius of a clump of mass of order $\Sigma H^2$. By choosing $G = 1 / \upi$, $\Omega = 1$ and the average density $\Sigma = 1$, we obtain $H = 1$ in code units. With this choice, the pressure scale-height, $H_{\rm p}$, is given by
\begin{equation}
H_{\rm p} = Q H,
\end{equation}
where $Q$ is Toomre's $Q$ parameter.

For the initial conditions we set up the profile in hydrostatic equilibrium with a vertically isothermal temperature profile, i.e. close the equilibrium temperature profile \citep{Shi2014}. The density structure is computed by solving for hydrostatic balance including both the external potential and the self-gravity assuming an infinite slab geometry. The initial temperature is set to give $Q=1$. We then add a perturbation to the $x$- and $y$-components of the velocity. Unless specified otherwise, following \citet{Johnson2003} the velocities were set according a Gaussian random field with a flat spectral distribution (white noise) in the range $0.25 < k < 4$ with an r.m.s. amplitude of $c_s$. In \autoref{Sec:Frag} we have also explored introducing uncorrelated velocities instead.

\section{Gravito-turbulence}

\subsection{Background properties}

\begin{figure*}
\begin{tabular}{cc}
\includegraphics[width=\columnwidth]{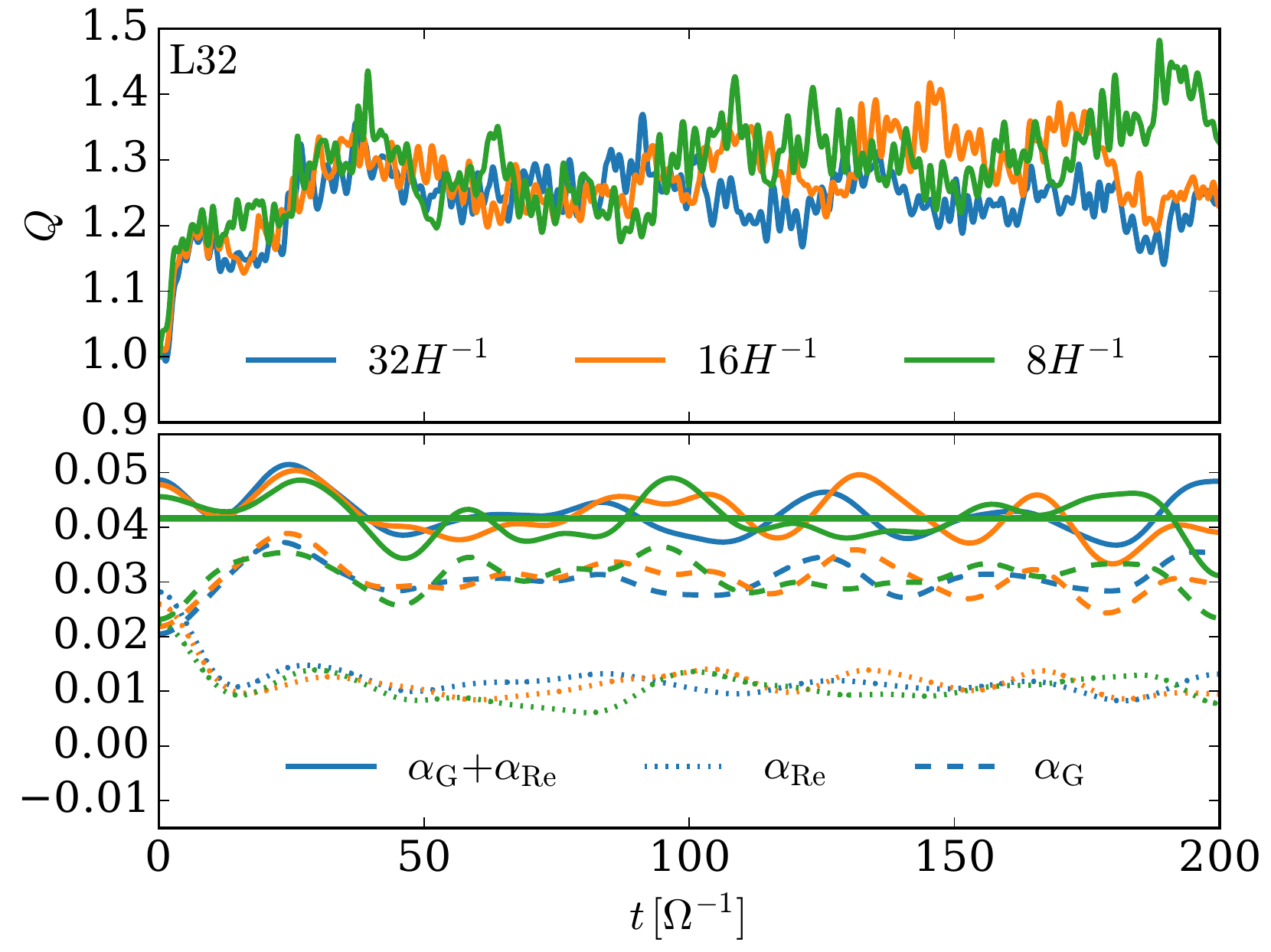}  & 
\includegraphics[width=\columnwidth]{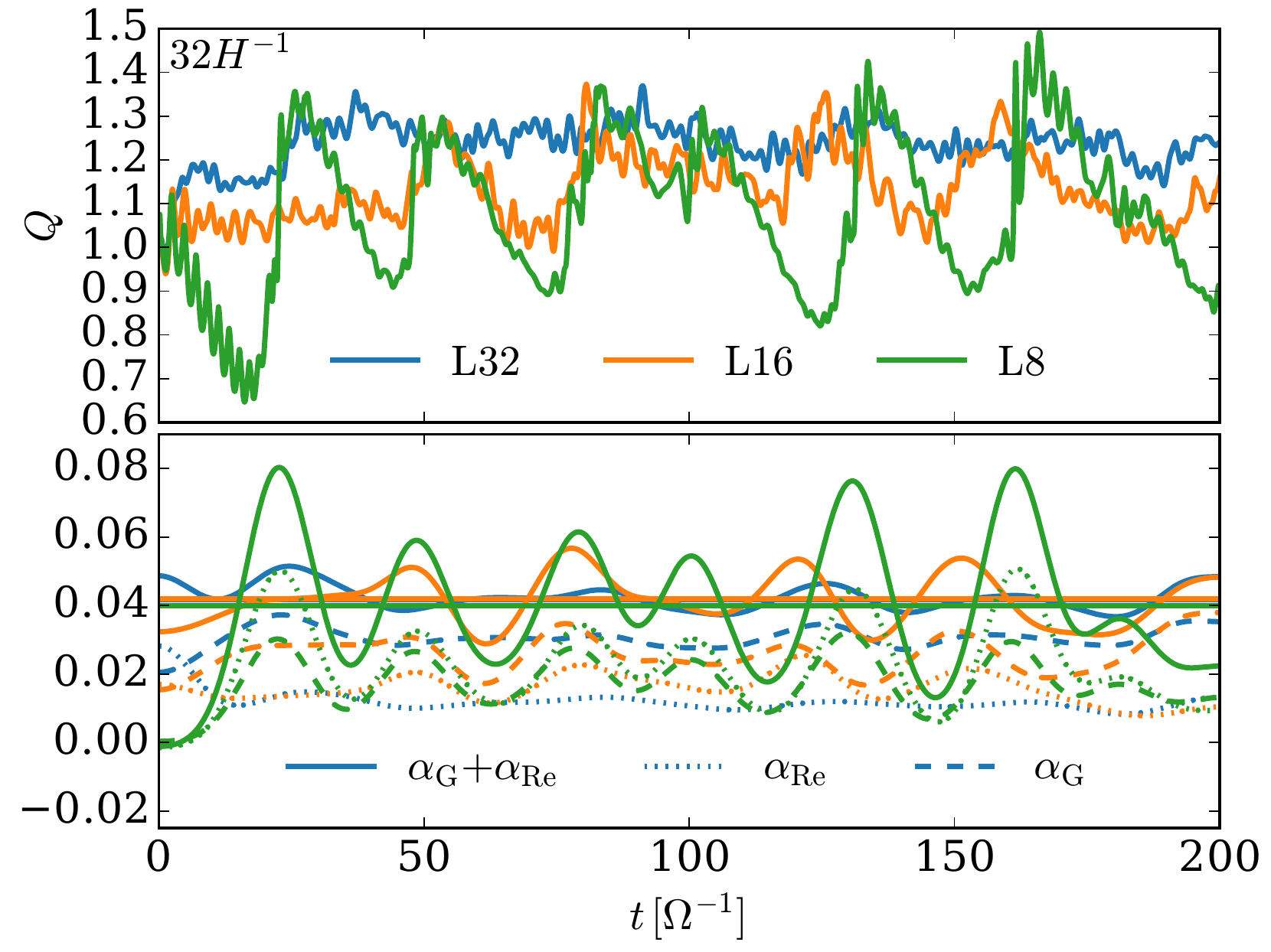} 
\end{tabular}
\caption{Dependence of volume averaged quantities on resolution (left) and box size (right). The Toomre $Q$ has been computed using the density-weighted r.m.s. sound speed. $\alpha_{\rm G}$ and $\alpha_{\rm Re}$ have been Gaussian smoothed over an orbital period to remove the high frequency oscillations. The solid horizontal lines show the time-averaged $\alpha$, which agree with \autoref{Eqn:GammieAlpha} to within a few per cent.}
\label{Fig:Qalpha}
\end{figure*}

\begin{figure*}
\includegraphics[width=\textwidth]{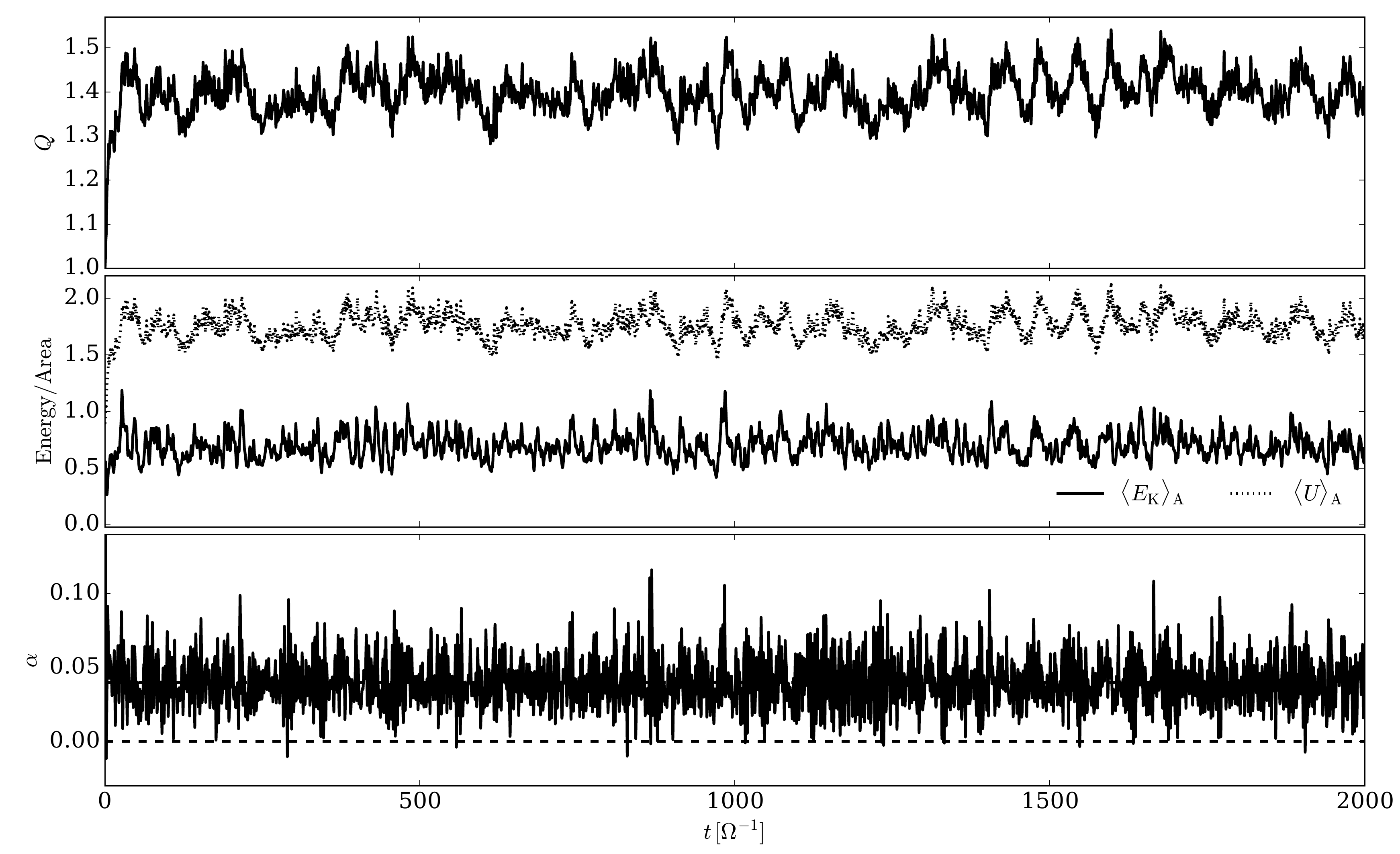} 
\caption{Evolution of the box-averaged quantities for the long term simulation, L64N8\_lt. Note that the range of $\alpha$ shown is much smaller than in \autoref{Fig:Ealpha}.}
\label{Fig:QEalpha_long}
\end{figure*}

\begin{figure*}
\includegraphics[width=\textwidth]{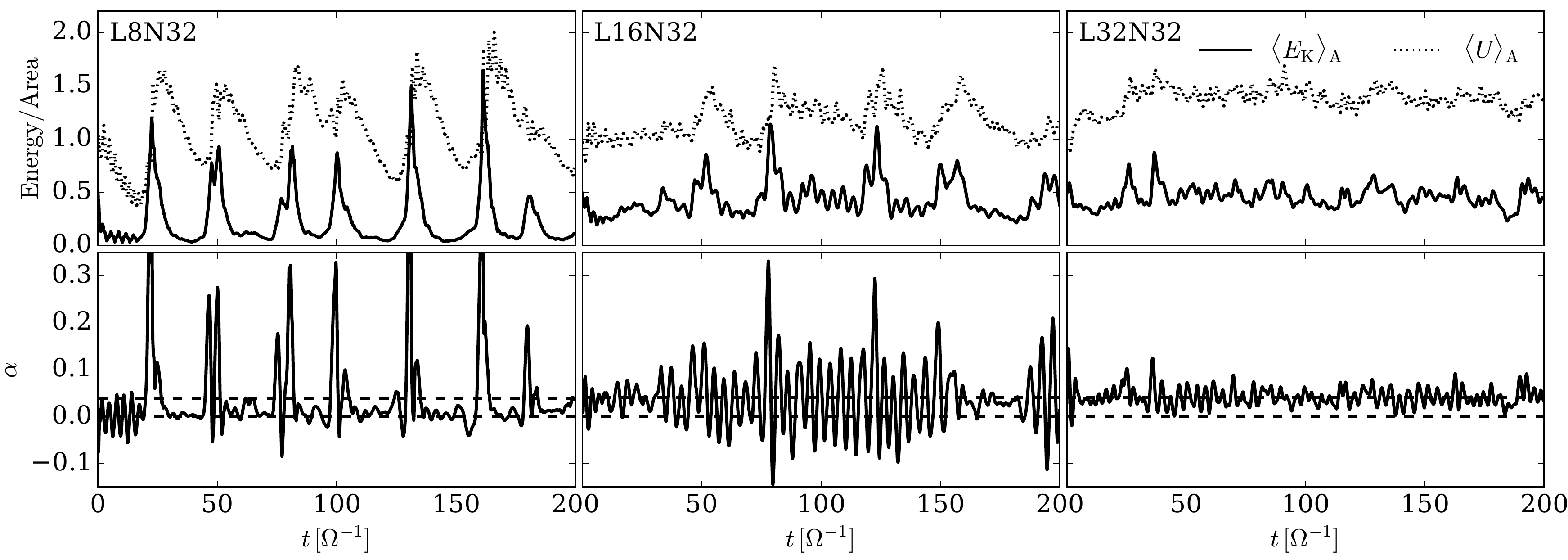} 
\caption{Top: Kinetic, $\langle E_{\rm} \rangle_{\rm A}$ and thermal $\langle U \rangle_{\rm A}$ energy per unit area for simulations with different box sizes. Bottom: The total stress $\alpha$. Unlike in \autoref{Fig:Qalpha} the stress has not been smoothed. Dashed lines show $\alpha = 0$ and $\alpha = 0.04$, the steady state value.}
\label{Fig:Ealpha}
\end{figure*}

\begin{table*}
\centering
\caption{Box averaged properties for the simulations with $\beta = 10$. Averages are computed from $t=50\Omega^{-1}$ to $t=200\Omega^{-1}$, except for the run L64N8\_lt, for which the range $50 \Omega^{-1}$ to $2000\Omega^{-1}$ is used.}
\label{Tab:SimStats}
\begin{tabular}{l ccc ccc c}
\hline
Name & Box Size ($L_x \times L_y \times L_z$) & Resolution ($N_x \times N_y \times N_z$) &
 $\langle Q \rangle $ & $\langle \alpha \rangle $ & $\langle \alpha_{\rm G} \rangle / \langle \alpha \rangle$ & $\langle  E_{K, x} + E_{K, y} \rangle_{\rm A} $ & $ \langle E_{K, z} \rangle_{\rm A}$ \\
\hline
L8N8  & $8H \times 8H \times 6H$ & $ 64 \times  64 \times  48$ & 1.13 & 0.039 & 0.49 & 0.20 & 0.032 \\
L8N16 & $8H \times 8H \times 6H$ & $128 \times 128 \times  96$ & 1.13 & 0.041 & 0.47 & 0.21 & 0.039 \\
L8N32 & $8H \times 8H \times 6H$ & $256 \times 128 \times 192$ & 1.11 & 0.040 & 0.44 & 0.20 & 0.041
\smallskip \\
L16N8  & $16H \times 16H \times 6H$ & $128 \times 128 \times  48$ & 1.19 & 0.041 & 0.64 & 0.42 & 0.042 \\
L16N16 & $16H \times 16H \times 6H$ & $256 \times 256 \times  96$ & 1.14 & 0.041 & 0.64 & 0.32 & 0.047 \\
L16N32 & $16H \times 16H \times 6H$ & $512 \times 512 \times 192$ & 1.15 & 0.042 & 0.61 & 0.41 & 0.051
\smallskip \\
L32N8  & $32H \times 32H \times 6H$ & $ 256 \times  256 \times  48$ & 1.30 & 0.042 & 0.76 & 0.44 & 0.064 \\
L32N16 & $32H \times 32H \times 6H$ & $ 512 \times  512 \times  96$ & 1.28 & 0.042 & 0.73 & 0.43 & 0.071 \\
L32N32 & $32H \times 32H \times 6H$ & $1024 \times 1024 \times 192$ & 1.24 & 0.042 & 0.73 & 0.39 & 0.070
\smallskip \\
L64N8  & $64H \times 64H \times 6H$ & $ 512  \times  512  \times  48$ & 1.42 & 0.039 & 0.86 & 0.64 & 0.080 \\
L64N8\_lt &                         &                                 & 1.40 & 0.040 & 0.86 & 0.62 & 0.076 \\
L64N16 & $64H \times 64H \times 6H$ & $ 1024 \times  1024 \times  96$ & 1.41 & 0.042 & 0.84 & 0.64 & 0.090
\smallskip \\
L128N8 & $128H \times 128H \times 6H$ & $ 1024 \times  1024 \times  48$ & 1.40 & 0.039 & 0.87 & 0.65 & 0.077 \\

\hline
\end{tabular}
\end{table*}

\begin{figure*}
\includegraphics[width=\textwidth]{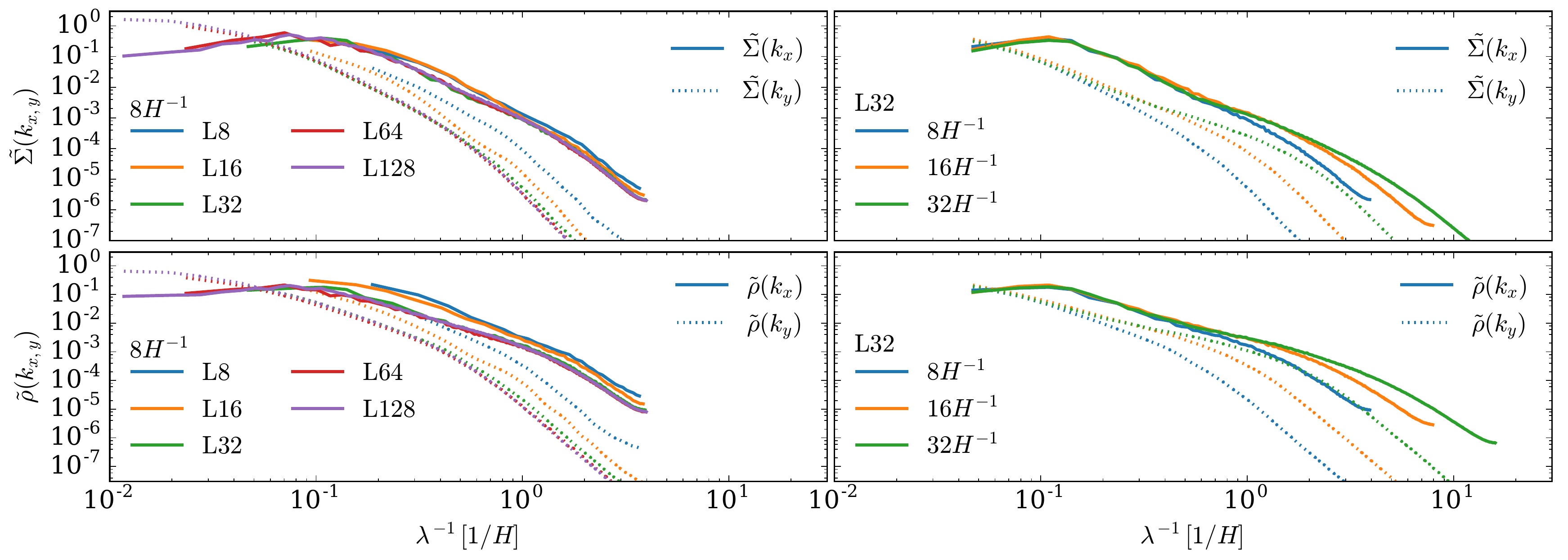} 
\caption{Power spectrum of density fluctuations for simulations with different box sizes (left) and resolution (right). Top panels show surface density fluctuations, while the bottom panels show the mid-plane volume density. The line-styles denote power spectra in the radial ($x$) and azimuthal ($y$) directions.}
\label{Fig:DensityPS}
\end{figure*}

The statistical properties of gravito-turbulence have been studied in great detail previously, here we briefly present their properties as a demonstration that the simulations are behaving as expected. For this purpose, we  explore how the properties depend on resolution and box size for a cooling time $\beta = 10$. \autoref{Fig:Qalpha} shows the box-averaged Toomre $Q$ parameter and stress, $\alpha$. Here we compute $Q$ via
\begin{equation}
Q = \frac{\Omega \langle c_s^2 \rangle_\rho^{1/2}}{\upi G \langle \Sigma \rangle}, \label{Eqn:Q}
\end{equation}
where $\langle X \rangle$ is the volume average,
\begin{equation}
\langle X \rangle = \frac{1}{V} \int_V X  \diff V,
\end{equation}
with $V = L_x L_y L_z$ and $\langle X \rangle_\rho$ is the mass-weighted volume average,
\begin{equation}
\langle X \rangle_\rho = \frac{\int_V \rho X \diff V}{\int_V \rho \diff V}.
\end{equation}
We further define the average per unit area: 
\begin{equation}
\langle X \rangle_{\rm A} = \frac{1}{A} \int_V X \diff V,
\end{equation}
where $A = L_x L_y$ such that $\langle \Sigma \rangle = \langle \rho \rangle_{\rm A}$. Here we note that the density-weighed r.m.s. sound speed used $\langle c_s^2 \rangle_\rho^{1/2}$ in the expression for $Q$ is typically a few per cent larger than the density-weighted sound speed, $\langle c_s \rangle_\rho$.

The stress $\alpha = \alpha_{\rm G} + \alpha_{\rm Re}$, where the gravitational and Reynolds stress-to-pressure ratios, $\alpha_{\rm G}$ and $\alpha_{\rm Re}$, are given by 
\begin{equation}
\alpha_{\rm G} = \frac{2}{3\gamma}\frac{\langle g_x g_y \rangle}{4 \upi G \langle P \rangle}.
\end{equation}
and 
\begin{equation}
\alpha_{\rm Re} = \frac{2}{3\gamma}\frac{\langle\rho v_x \delta v_y\rangle}{\langle P \rangle},
\end{equation}
Here $\delta v_y = v_y + q \Omega x$ and $g_x$, $g_y$ are the x- and y-components of the gravitational acceleration respectively.

From \autoref{Fig:Qalpha} we see that neither $Q$ or $\alpha$ are sensitive to the resolution of the simulations suggesting that our lowest resolution is already sufficient to capture the large-scale behaviour of the gravito-turbulence. This is consistent with previous results, which also found convergence at resolutions around 8 cells per scale-height \citep{Shi2014,Riols2017}. We note that the convergence of the total $\alpha$ is not a particularly stringent test of the code because $\alpha$ is fixed by energy conservation in a quasi-steady gravito-turbulent equilibrium, as long as the energy lost through the boundaries is small (which we estimate to be a few per cent per cooling time). The balance between the heating and cooling gives
\begin{equation}
\alpha = \frac{4}{9\gamma (\gamma-1) \beta} \label{Eqn:GammieAlpha}
\end{equation}
\citep{Gammie2001}, hence $\alpha = 0.04$ for $\beta = 10$, in excellent agreement with the simulations at all resolutions and box sizes. However, the independence of $Q$ and the ratio of gravitational stress to total stress with resolution confirms that the large-scale structure is well resolved. 

Conversely, there is a box-size dependence for $L \lesssim 64H$. For small boxes the gravitational stress increases with box size, while the Reynolds stress decreases to maintain the total stress \citep[see also][]{Riols2017}. Similarly, the kinetic energy increases with box size. However, we do see convergence for our largest box sizes, with the averaged statistics agreeing well for boxes with sizes $64H$ and $128H$ (\autoref{Tab:SimStats}), in agreement with the results of \citet{Shi2014} at lower resolution. While these averages have only been computed over $150\Omega^{-1}$, our simulations do not show clear long term trends. This has been confirmed by running the low resolution $64H$ model until $2000\Omega^{-1}$, which shows near identical averages to those obtained between $50\Omega^{-1}$ and $200\Omega^{-1}$. 

For large domains, the global quantities are remarkably steady, showing no long term trends in $Q$, $\alpha$, or the averaged kinetic and thermal energies (\autoref{Fig:QEalpha_long}). Similar results were seen in the 2D simulations of \citet{Gammie2001}, which also employed large domains. These results are contrary to those reported by \citet{Riols2017}, who found trends on time-scales of 100s of orbits in simulations with a domain size of $20H$. \citet{Riols2017}, attribute this to transient density clumps seen in their simulations. We argue that this behaviour is affected by the small box-size, which gives rise to unsteady behaviour as described below. 

The smallest domains show a different behaviour, characterized by bursts on time-scales regulated by the cooling. Instead of the slow variation seen for $L_x \ge 32H$, in small boxes the Toomre $Q$ parameter undergoes phases of rapid increase followed by steady decrease on the cooling time-scale. This behaviour is clearest for $L_x = 8H$, but can also be seen for $L_x = 16H$. To elucidate the origin of this bursty behaviour, we show the kinetic energy and total stress, unsmoothed, together in \autoref{Fig:Ealpha}. The smallest boxes undergo rapid bursts that heat the disc, after which kinetic energy and stress rapidly decay. With the heating removed the disc cools until reaching $Q \approx 1$, before undergoing another burst and so the process repeats. We note that this bursty behaviour can also be seen in the temporal variations of $Q$ and $\alpha$ in the simulations presented by \citet{Baehr2017}.

To provide an explanation for the transition to bursty behaviour as the box-sizes decreases, we consider how the power spectrum of density fluctuations varies with box-size.

To compute the power spectra, we first compute the shear-periodic Fourier transform, $\hat{X}(k_x, k_y, t)$, following the method outlined in \citet{Hawley1995}\footnote{The wavenumbers, $k_x$ and $k_y$, are the Eulerian wavenumbers in Hawley's terminology.}. Denoting the time-average of $\hat{X}(k_x, k_y, t)$ as $\hat{X}(k_x, k_y)$ (over the temporal range $t=50$ to $200\Omega^{-1}$), the 1D and 2D power spectra are then computed via
\begin{equation}
\tilde{X}(k_x) = \int |\hat{X}(k_x, k_y)|^2 \diff k_y
\end{equation}
and
\begin{equation}
\tilde{X}(k) = \int |\hat{X}(k_x, k_y)|^2 \delta (k_x^2 + k_y^2 - k^2) \diff k_y \diff k_y,
\end{equation}
where $\delta(x)$ is the Dirac $\delta$-function. 

From \autoref{Fig:DensityPS} we see that the power spectra with $L \gtrsim 32\,H$ are in good agreement, whereas smaller boxes show larger amplitude fluctuations on all scales present. However, we note that while the power spectra for $L = 32 H$ are in excellent agreement with larger boxes on small scales, the wavelength of the peak mode (7 to $10\,H$) is smaller than that seen in larger boxes ($14H$). While in both cases the peak radial wavelength is considerably smaller than the box size, the peak azimuthal wavelength is always close to the box-size. This makes it challenging to obtain direct measurements of the dominant azimuthal wavelength from the power-spectrum because the separation between $k$ points is comparable to the $k$ itself. Thus we find that the pitch angle of the spirals provides a more robust constraint on the $k_y$ where the density power spectrum peaks in any given simulation. The pitch angles are measured from the 2D auto-correlations computed from  the time-averages, $|\tilde{\rho}(k_x, k_y)|^2$ and $|\tilde{\Sigma}(k_x, k_y)|^2$, via inverse Fourier transform \citep[see, e.g.][]{Gammie2001}. Fitting for the longest axis in the auto-correlation function produces a consistent estimate of the pitch angle for all box sizes, $i \approx 13^\circ$, from which we can estimate $k_y = k_x \tan i$. For $L = 64\,H$ and $128\,H$, we find $\lambda_y = 2\upi / k_y \approx 60H$, while for boxes with $L = 32H$ and smaller this estimate for $\lambda_y$ decreases, remaining close to the box-size. Thus we argue that the squeezing of azimuthal wavelengths by small boxes is the driver for the non-convergence.

Since the squeezing of the azimuthal wavelengths in small boxes is significant, when considering
protostellar discs  (which have $H_{\rm p}/R \approx 0.1$) the required domain size becomes  comparable to the size of the disc. This means that the local model is not fully applicable, and thus the precise properties of
gravito-turbulence will differ from those determined here, even before additional physics such as the radiative transfer are factored in.

For the smallest boxes, $L \lesssim 16H$, the bursty nature is then understood as being due to the fact that the modes otherwise responsible for maintaining $Q\approx 1$ are larger than the box size, and thus not present in the simulation. Since the small wavelength modes are only unstable at lower $Q$ \citep{Mamatsashvili2010}, this explains the on average lower $Q$ in the boxes. However, we see from \autoref{Fig:DensityPS} that the density perturbations have a larger amplitude even on average, and from \autoref{Fig:Ealpha} that they give rise to rapid heating when active. This drives the box to high $Q$ where all of the modes with wavelength smaller than the box size are again stable.  This behaviour can also be clearly seen in the profiles of $Q$ and $\alpha$ for the box sizes of $12\,H$ presented by \citet{Baehr2017}. 

While this effect is likely present in a weaker form at intermediate box sizes, the unsteady behaviour seen in \citet{Riols2017} is also affected to the formation of transient clumps, which significantly perturb the boxes. For small boxes, the periodic nature of the domain may inflate their importance as the clumps self-interact, an effect that should be weaker in larger domains. Once the perturbation is present, the large amplitude modes also persist for much longer than the clumps themselves. This may partially be another consequence of the perturbation being limited to a small number of modes.

This insight (concerning the importance of long wavelength modes in allowing a steady gravito-turbulent state to develop) also provides an explanation for a hitherto unexplained phenomenon. \citet{Lodato2005} showed in global simulations that there is no steady self-gravitating state once the disc-to-star mass ratio exceeds around 0.25, at which point $H/R \sim 0.1$. Our simulations demonstrate bursty behaviour if the azimuthal domain is less than $\sim 60 H$, which corresponds (at this $H/R$ value) to $\sim 2 \pi R$. We therefore suggest that the onset of bursty behaviour in global simulations at high disc to star mass ratio is also a consequence of the missing long wavelength modes (but where this is not numerical, as in the present experiments, but related to the finite disc size). We note that this comparison can only be qualitative in detail due to differences between the shearing box model and real discs, for example through the radial periodic boundaries or the neglection of curvature terms.

\subsection{Small-scale structure and transition to three-dimensional dynamics}

\begin{figure*}
\includegraphics[width=\textwidth]{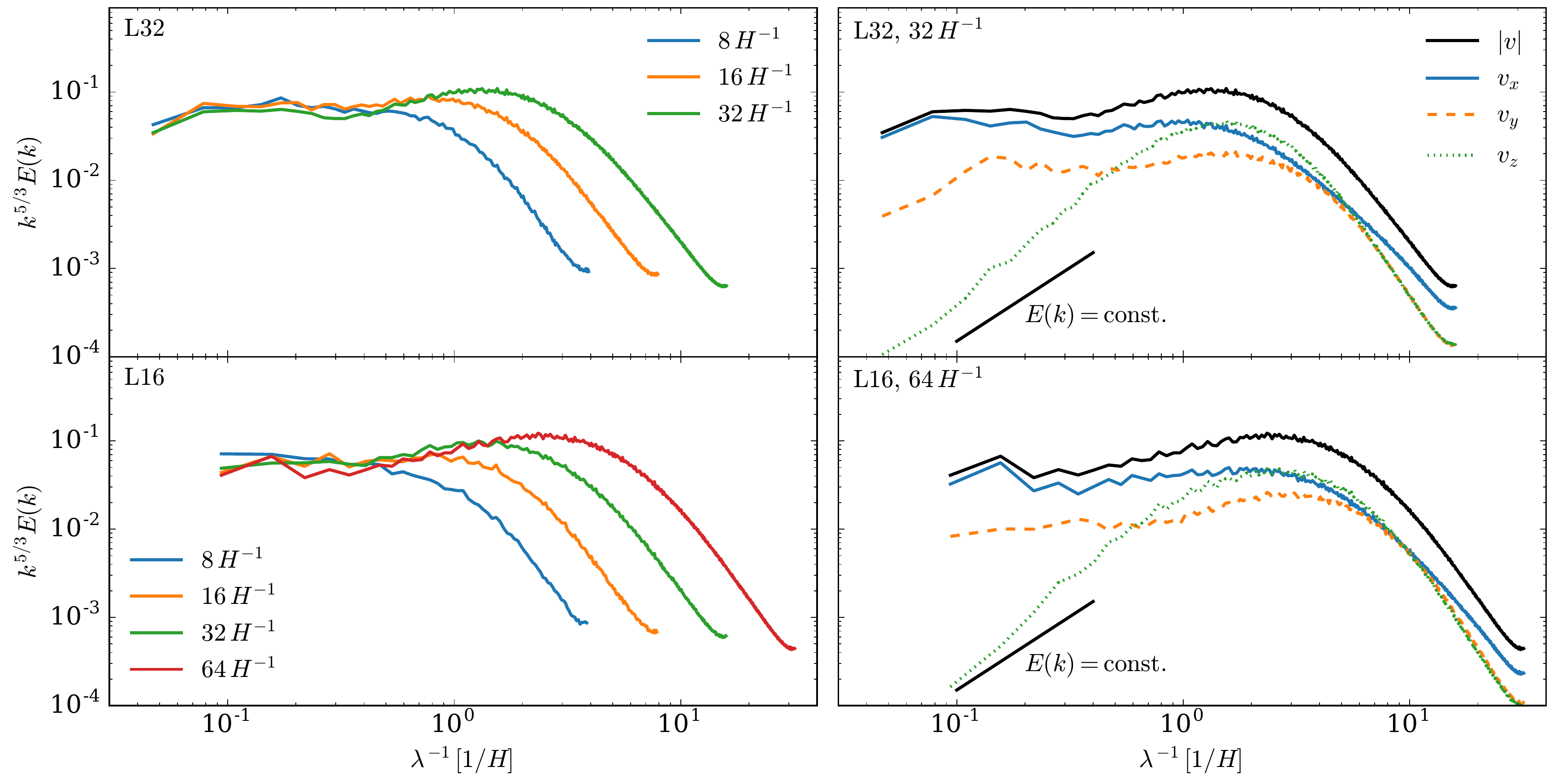}
\caption{Mid-plane kinetic energy spectrum weighted by $\rho^{1/3}$ and scaled by $k^{5/3}$ vs inverse wavelength, $\lambda^{-1}$. Left: variation of the power spectrum of the total velocity with resolution. Right: The contribution to the total power from each velocity component in the high resolution spectrum. Top panels show simulations with box sizes of $L = 32\,H$, while bottom panels show $L=16\,H$.}
\label{Fig:Kolmogorov}
\end{figure*}

\begin{figure*}
\includegraphics[width=\textwidth]{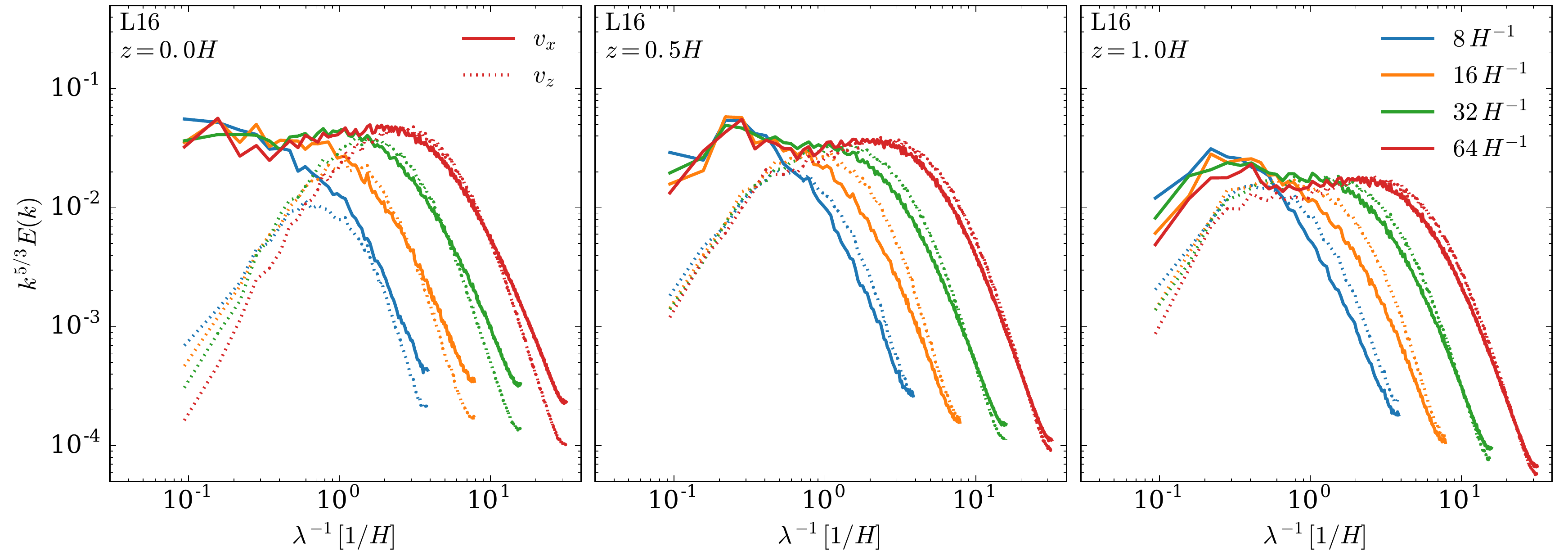}
\caption{Weighted kinetic energy power spectra for the $x$- and $z$-components of the velocity for different vertical slices.}
\label{Fig:TurbZ}
\end{figure*}

The small-scale properties of three-dimensional gravito-turbulence have been characterized by \citet{Bournaud2010} in the barotropic limit (short cooling time) and at moderate cooling times by \citet{Riols2017}. We begin by summarizing the key results of these studies. 

\citet{Bournaud2010} showed that power spectra of the in plane ($v_x$ and $v_y$) components of the motions approximately follow the $k^{-5/3}$ scaling expected for both two- and three-dimensional incompressible turbulence at all scales (neglecting rotation), with the vertical component ($v_z$) also following a similar scaling for wavelengths shorter than the scale-height, $H$, but falling off at larger scales. They also showed that the spectrum of enstrophy, $|\nabla \times v|^2$, followed the $k^{-1}$ behaviour on large scales that is expected for two-dimensional turbulence \citep{Kraichnan1967}, thus arguing that gravito-turbulence transitions from two to three-dimensional character at these scales. Furthermore, they found that this is associated with a break in the power spectrum of density fluctuations consistent with that seen in the Large Magellanic Cloud \citep{Elmegreen2001}.

\cite{Riols2017} also found that the kinetic energy spectrum ($\rho^{1/3}$-weighted, see below) followed $k^{-5/3}$ on large scales, and identified a parametric instability of the large-scale modes by inertial waves as a possible origin of the small-scale turbulence. The incompressible nature of the inertial waves may help to explain why the kinetic energy power spectrum of gravito-turbulence is close to that expected for incompressible turbulence. Although \cite{Riols2017} found the power spectrum steepened on scales smaller than the scale height, we show here that this is likely due to the numerical resolution. We will thus argue that the picture put forward by \cite{Bournaud2010} largely applies to protostellar gravito-turbulence.

In addition to the density power spectrum, we consider the power spectrum of $\vec{u} = \rho^{1/3} \vec{v}$. The $\rho^{1/3}$ weighting is included to account for the effects of compressibility, with the power spectrum of $w$, which denote as $E(k)$, following $k^{-5/3}$ as long as the simple scaling arguments still hold \citep{Lighthill1955,Fleck1996,Kritsuk2007}. Again, we consider the two-dimensional spectra evaluated at a particular height above the mid-plane. 

\autoref{Fig:Kolmogorov} shows the mid-plane kinetic energy spectrum for the total velocity $|v|$ and individual components. These have been averaged over $150\Omega^{-1}$, except for the simulation with $L = 16$ and resolution $64\,H^{-1}$, which was averaged over a shorter run of $75\Omega^{-1}$. First, we note that numerical diffusion affects the power spectra on scales well above the grid scale, causing significant dissipation on length scales smaller than about 10 to 20 cells. Given that \citet{Riols2017} used a resolution of around $20H^{-1}$, this potentially explains the cut off at $\sim H$, although the exact scale likely depends on the details of the numerical method.

Here we use the 1D power spectrum in $k$, but note that the same behaviour is seen in $k_x$ and $k_y$ independently. The main difference between spectra in $k_x$ and $k_y$ occurs at large scales, with $k_y$ spectrum peaking at longer wavelengths than $k_x$. These peaks occur at comparable scales to the peaks in the density power-spectrum seen in \autoref{Fig:DensityPS}, confirming those scales as the upper limit to the scales of gravito-turbulence.

On large scales the power spectrum agrees closely with $k^{-5/3}$; hover, as the wavelength drops below $H$ the total kinetic energy is less steep until turning over due to dissipation. Examining the individual components, we see that this is driven primarily by the increasing contribution from $v_z$. The spectra of $v_x$ and $v_y$ are instead close to $k^{-5/3}$ until dissipation becomes significant. 

We note that the mid-plane power spectrum of $v_z$ peaks close to where the power spectra of $v_x$ and $v_z$ are comparable, which, for horizontal slices near the mid-plane, is close to the dissipation scale at all resolutions (\autoref{Fig:TurbZ}). While it may be possible that the peak of $v_z$ will always remain close to the dissipation scale, we argue that this is unlikely, instead preferring the explanation that the simulations are not yet sufficiently resolved (except perhaps the highest resolution simulation). To justify this, we also present the power spectra of $v_x$ and $v_z$ for slices at $z=0.5H$ and $z = H$, where the flow is more isotropic  \citep{Riols2017}. From \autoref{Fig:TurbZ} we see that there is a $z$-dependent   above which the power-spectra of $v_x$ and $v_z$ become equal, and that this scale is not affected by resolution. Between this scale and the dissipation scale, the power-spectra of $v_x$ and $v_z$ are both in good agreement with $k^{-5/3}$. 

\autoref{Fig:TurbZ} also suggests that at small scales the flow becomes isotropic. While this is clear for $v_x$ and $v_z$, the power spectrum of $v_y$ remains smaller $v_x$ at all scales and resolutions. However, the difference decreases towards smaller scales, suggesting that at small enough scales the flow becomes isotropic. Extrapolating the power spectra suggests that the flow may become fully isotropic below $\lambda \sim 0.1H$. 

The fact that the small-scale flow only becomes isotropic for scales $\lambda \lesssim 0.1$ to $0.3H$ may help to explain why we do not see a break in the power spectrum of density fluctuation. \citet{Bournaud2010} only see such a break at the scale where the power-spectra of $v_x$ and $v_z$ are comparable, which is close to the dissipation scale in our simulations. Thus, the break in the density power spectrum and its slope scales smaller than this may be hidden by dissipation.  

\begin{figure}
\includegraphics[width=\columnwidth]{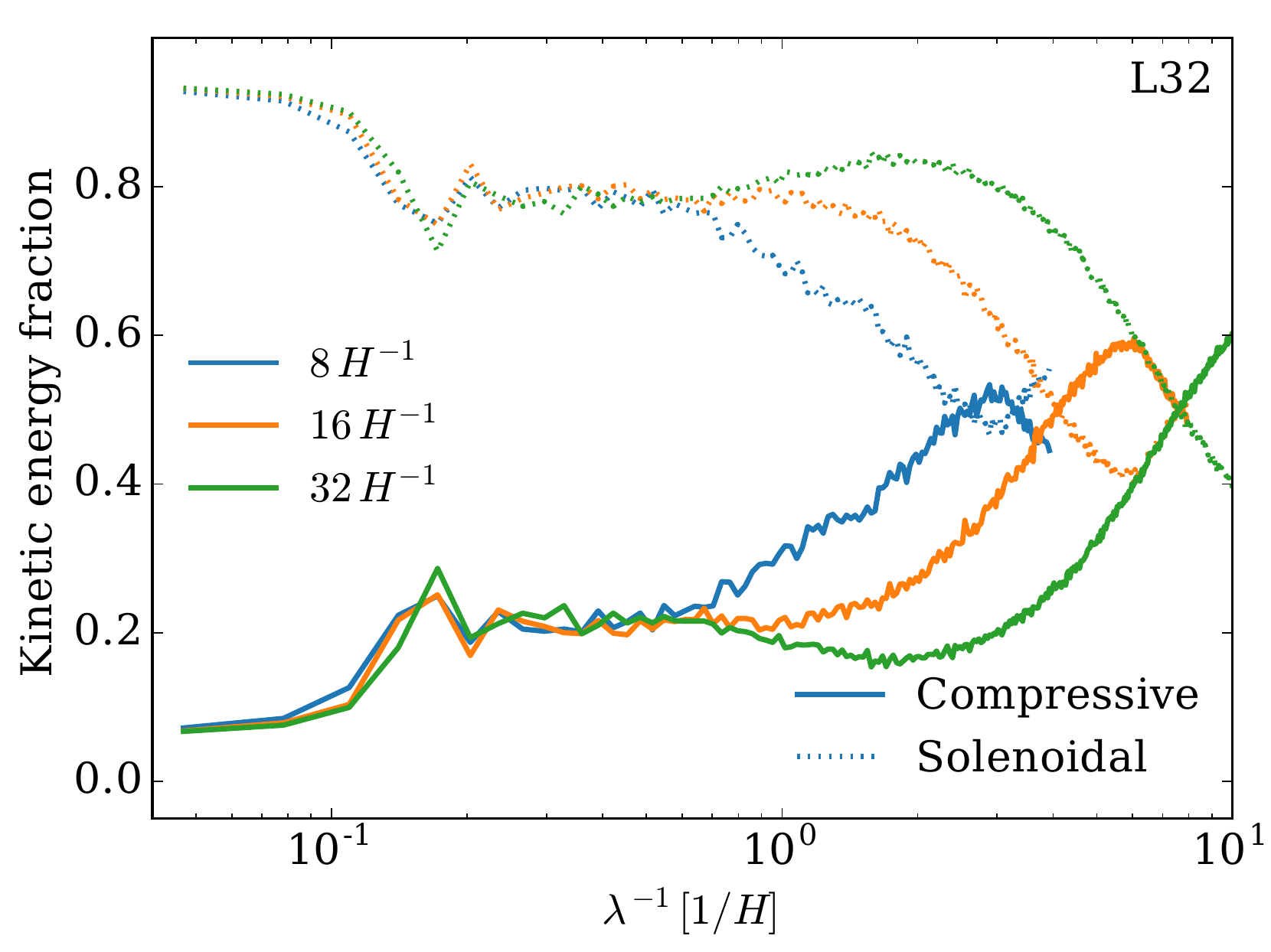}
\caption{Fraction of the density-weighted kinetic energy associated with solenoidal and compressive modes at different scales.}
\label{Fig:Solenoid}
\end{figure}

The agreement of the weighted kinetic energy spectra with $k^{-5/3}$ is further evidence that the small scale turbulence is not driven by large scale compressive motions associated with the spiral arms. \citet{Federrath2013} showed that the spectra can be steeper than 
$k^{-5/3}$ in the presence of strongly compressive forcing; however, if anything our simulations show spectra that may be slightly shallower than $k^{-5/3}$. By partitioning $\rho^{1/3}\vec{v}$ into compressive and solenoidal (incompressible) motions, we see that solenoidal motions dominate away from the grid scale (\autoref{Fig:Solenoid}). If compressive motions were important, they should dominate on the scales where the turbulence is driven. This explains the good agreement with  $k^{-5/3}$ and is consistent with the turbulence being driven by the inertial modes (that have incompressible character) as identified by \citet{Riols2017}. Thus we argue that gravito-turbulence really is turbulent on scales below the most unstable wavelength, $k H \approx 1$. At large scales the turbulence has two-dimensional character and transitions to three-dimensional character for $k H \gtrsim 20$ ($\lambda \lesssim 0.3 H$).

\subsection{Temporal correlations and diffusion}
\label{Sec:Diffuse}

\begin{figure*}
\includegraphics[width=\textwidth]{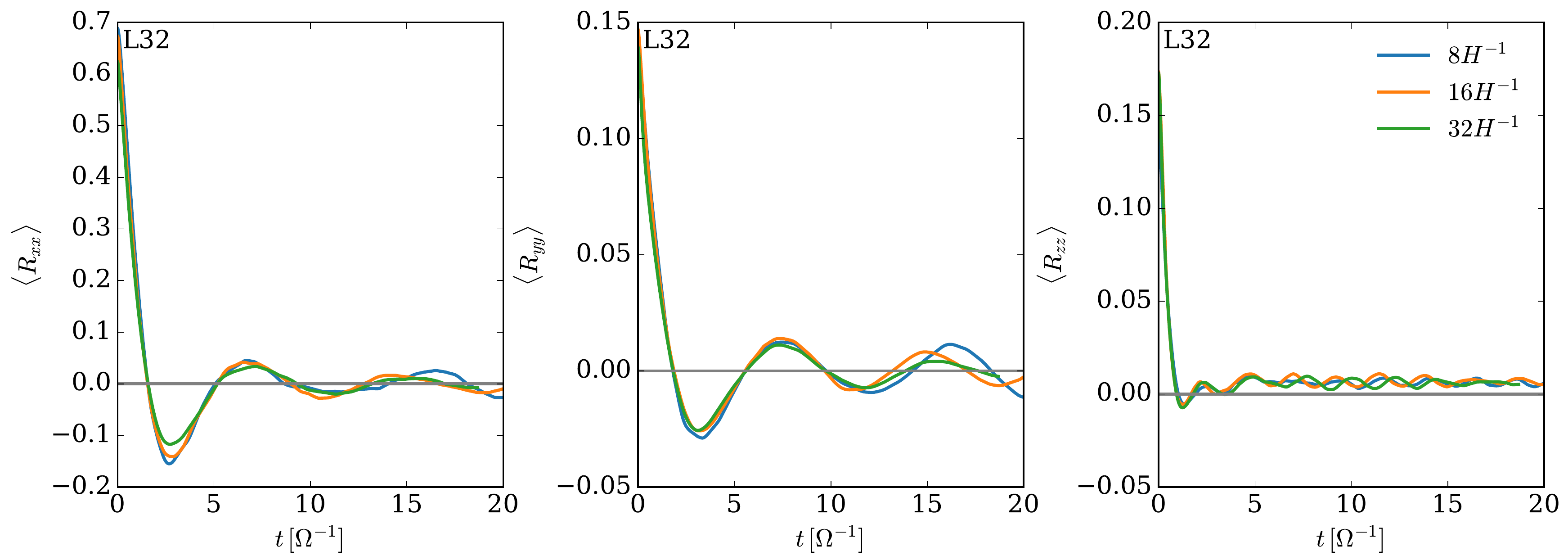}
\includegraphics[width=\textwidth]{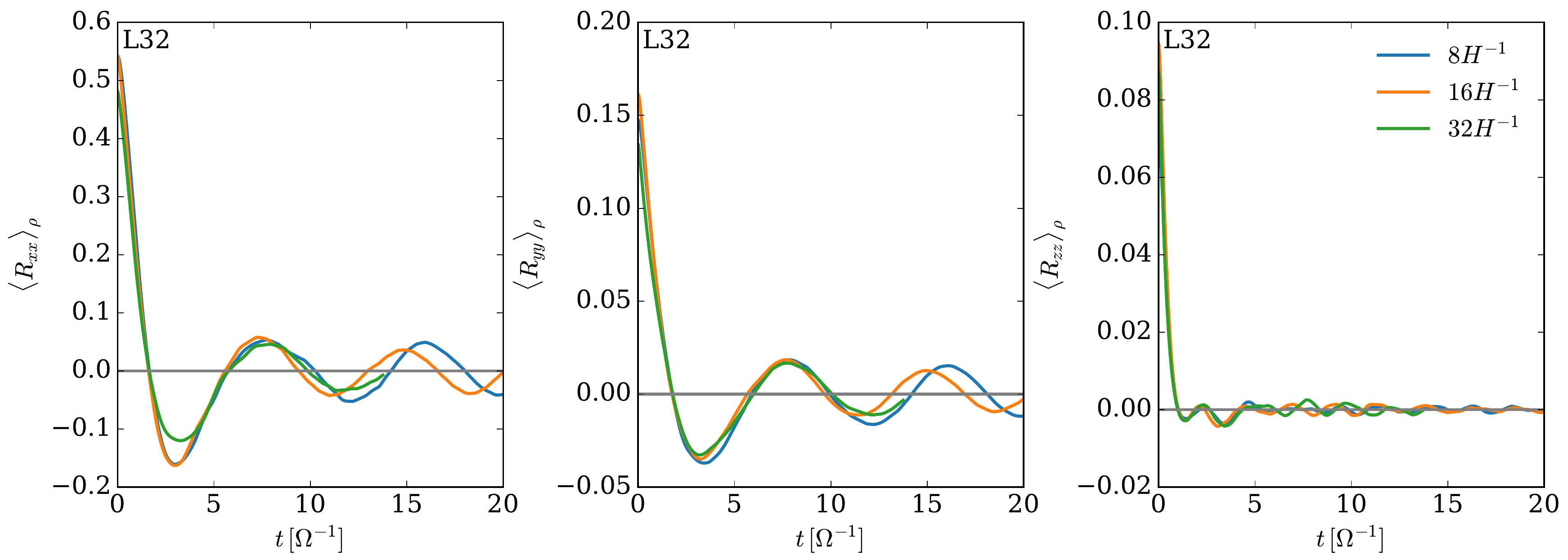}

\caption{Top: Temporal correlations functions of the velocity components. Bottom: Same, but density-weighted.}
\label{Fig:Rxx}
\end{figure*}

The radial diffusion coefficient for \textbf{gas or well-coupled particles}, $D_{x}$, is computed via
\begin{equation}
D_{x} = \int \langle R_{xx}(t) \rangle \diff t,
\end{equation}
where $\langle R_{xx}(t) \rangle$ is the correlation function of the Lagrangian velocity along trajectories,
\begin{equation}
\langle R_{xx}(t) \rangle = \langle v_x(t) v_x(0) \rangle.
\end{equation}
Here, we follow \citet{Zhu2015} and use the shear-corrected Eulerian velocity in the place of the Lagrangian velocity, averaging this over space and time. 

\citet{Shi2016} noted that in 2D gravito-turbulent boxes, the diffusion coefficients as computed above are dominated by low density regions between the spiral arms, where the velocities are large. They showed that the measured diffusion of solid particles is in better agreement with a mass-weighted diffusion coefficient, which only moderate differences between the coefficients computed using the dust and gas density. Thus we also compute, 
\begin{equation}
\langle R_{xx}(t)  \rangle_\rho = \frac{\langle  \rho(t) v_x(t) \rho(0) v_x(0) \rangle}{ \langle  \rho(t) \rho(0) \rangle},
\end{equation}
along with the equivalent diffusion coefficient.

In \autoref{Fig:Rxx} we show the temporal correlation functions. The radial and azimuthal correlation functions,  $\langle R_{xx}(t) \rangle$ and $\langle R_{yy}(t) \rangle$, show large oscillations with an angular frequency $\omega \approx 0.8 \Omega$, which dominates the correlation function. This is likely driven by a large-scale epicyclic mode with the same frequency that dominates the dynamics \citep{Riols2017}. We note that this oscillation makes estimating the diffusion coefficient from $\langle R_{xx}(t) \rangle$ and $\langle R_{yy}(t) \rangle$ challenging because the oscillations largely cancel out. The vertical correlation, $\langle R_{zz}(t) \rangle$, does not show this oscillation.

We find that while $\langle R_{zz}(t) \rangle$ decays on a time-scale much shorter than the dynamical time, there is a small background component that remains positive on much longer time-scales, which would appear to suggest a small net inflow or outflow. However, this is not seen in density-weighted coefficient, $\langle R_{zz}(t) \rangle_\rho$, and is driven by the low density regions above the mid-plane, which may be affected by the boundary conditions. This velocity residual affects the estimate of the vertical diffusion coefficient; however, once subtracted off this results in estimates of $D_{z}$ (0.02 to 0.05) that are similar between the mass-weighted one (0.025 to 0.03). Comparing this to the gravito-turbulent $\alpha=0.04$ gives an estimate of the Schmidt number of 1 to 2 (the ratio of turbulent angular momentum transport and mass diffusion coefficients), similar to the in plane diffusion coefficients found in 2D \citep{Shi2016}.

\section{Fragmentation}
\label{Sec:Frag}

\subsection{Prompt fragmentation}

It is well known that \emph{global} simulations of gravito-turbulent discs may undergo spurious fragmentation before gravito-turbulence becomes properly developed \citep{Paardekooper2011,Young2015,Deng2017}. These works found that fragmentation can be spuriously triggered at interfaces between turbulent and non-turbulent regions that arise due to radial variations in the cooling time. Here we report that shearing box simulations started from dynamically cold initial conditions can also undergo a similar prompt fragmentation before the gravito-turbulence develops properly.

\begin{figure}
\includegraphics[width=\columnwidth]{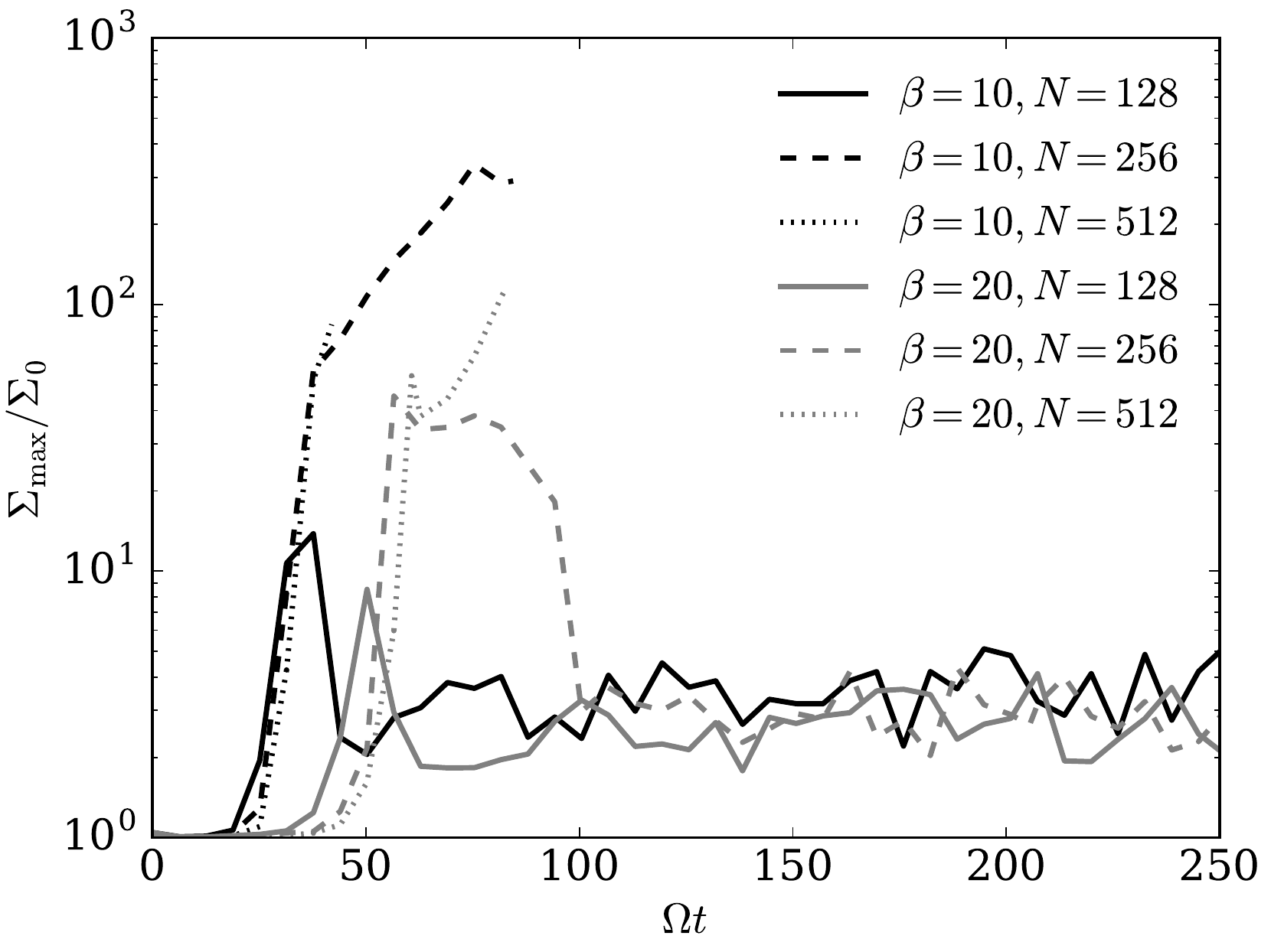}
\caption{Maximum surface density as function of time for different cooling times and resolution. A peak in the surface density occurs just before the onset of gravito-turbulence. This can lead to a resolution dependence in the fragmentation, with higher resolution simulations achieving higher densities and thus being able to fragment at higher $\beta$. At $\beta  = 50$ (not shown) none of the models fragment. We also do not observe fragmentation in any of the models that become fully gravito-turbulent even after integrating for $200\upi\Omega^{-1}$.}
\label{Fig:Prompt}
\end{figure}

An example of this is shown in \autoref{Fig:Prompt}, where the surface peak density as a function of time is shown for simulations run at different $\beta$ and resolution with a fixed box size of $16H$. The velocities of each cell were initialized to independent random uniform values with an amplitude of $0.1c_s$. Here we see that at low resolution the maximum surface density rises to a high value before settling down to the $\sim 3\Sigma_0$ indicative of gravito-turbulence. However, at high enough resolution and low enough $\beta$ fragmentation can occur when the density perturbation reaches its maximum. This results in a resolution dependent fragmentation threshold ($\sim 16H^{-1}$ at $\beta=10$, $\sim 32H^{-1}$ at $\beta=20$, we did not see fragmentation at all at $\beta = 50$), and one that likely also depends sensitively on the numerical method.

The non-convergence of the previous results may not be entirely surprising because the gravitational instability has a preferred wavelength, $kH \approx 1$, while purely random velocities imply the perturbations have an effective length scale that varies with the grid scale of the simulation. However, we note that similar experiments using our default initial conditions, where the velocities are initialized according to a Gaussian random field. Even when the same $k$-space realization was used at all resolutions, we still found a resolution dependence on the fragmentation threshold, with the highest resolution simulation fragmenting at $\beta=10$ and an initial velocity amplitude of $0.1c_s$. By increasing the amplitude of the perturbation to $c_s$ none of the simulations tested fragmented (up to the maximum resolution of $32H^{-1}$).

Rather than attempt to find a set of initial conditions that show convergence, we avoid the problem in tests below by first relaxing the boxes into equilibrium at longer cooling time. This process has typically shown convergence in global simulations, and we demonstrate below the same result here.

\subsection{Relaxed initial conditions}
\label{Sec:dBdt}

\begin{table}
\caption{The cooling time, $\beta_{\rm frag}\Omega^{-1}$, at which a fragment first forms in the experiment where the cooling time was steadilty reduced. The simulations were initially relaxed into a gravito-turbulent state and $\beta$ was reduced linearly such $\beta(t) = \beta_0  - t / \delta t$. The box size was $16 H$.}
\label{Tab:Frag}
\begin{tabular}{lccc}
\hline
Resolution &  $\beta_0$ & $\delta t$ &  $\beta_{\rm frag}$ \\
(cells per $H$) & & $(\Omega^{-1})$ &\\
\hline
8 & 50 & $ 4\upi$  & 2.0 \\
8 & 50 & $ 8\upi$  & 2.0 \\
8 & 20 & $ 16\upi$ & 2.1 \\
\rule{0pt}{3.5ex}16 & 50 & $ 4\upi$ & 2.0 \\
16 & 50& $ 8\upi$  & 4.0 \\
16 & 20& $ 16\upi$ & 4.5 \\
\rule{0pt}{3.5ex}32 & 50 & $ 4\upi$ & 1.0 \\
32 & 50 & $ 8\upi$  & 5.0 \\
32 & 20 & $ 16\upi$ & 4.25 \\
\hline
\end{tabular}
\end{table}

Here we consider the stability of self-gravitating discs against fragmentation beginning from states that are already in gravito-turbulent equilibrium. We consider two experiments to determine the fragmentation threshold. First, we continually reduce the cooling time until the disc eventually fragments. Second we instantaneously reduce the cooling time to a new value and then investigate whether the disc fragments.

In the first experiment, we slowly reduce $\beta$ according to $\beta (t) =  \beta_0 - t / \delta t.$ The behaviour of the disc can be divided into two cases depending on whether $\delta t$ is smaller or greater than the cooling time. Using 3D global SPH simulations \citet{Clarke2007} showed that for fast changes in $\beta$ (small $\delta t$) the disc is not able to respond thermally to changes in $\beta$ and fragmentation is delayed until smaller $\beta$. However, for slow changes in $\beta$ the fragmentation boundary converged to $\beta \approx 3$, similar to what \citet{Gammie2001} found the stability criterion to be for 2D shearing-sheets.

\begin{table}
\caption{Whether a given simulation fragmented on the cooling time scale (Yes) or fragmented stochastically sometime after the new gravito-turbulent equilibrium was obtained, in which case we show the approximate number of dynamical times before fragmentation. Blank means the simulation did not fragment within $100\Omega^{-1}$.}
\label{Tab:FragDirect}
\begin{tabular}{l ccccc}
\hline
\multirow{2}{*}{Name} & \multicolumn{5}{c}{$\beta$} \\
 & 1 & 2 & 3 & 4 & 5 \\
\hline
L8N8   & Yes & Yes & Yes &  &  \\
L8N16  & Yes & Yes & Yes &  &  \\
L8N32  & Yes & Yes & Yes & Yes &  
\smallskip \\
L16N8  & Yes & Yes & Yes &  &  \\
L16N16 & Yes & Yes & Yes & 20 &  \\
L16N32 & Yes & Yes & Yes & 10 & 
\smallskip \\
L32N8  & Yes & Yes & 50 &  &  \\
L32N16 & Yes & Yes & Yes &  &  \\
L32N32 & Yes & Yes & Yes & 20 &  \\
\hline
\end{tabular}
\end{table}

We limit our exploration of the response to reducing the cooling time to box sizes of $16H$ due to the long run time needed for these experiments. For this test to produce a meaningful $\beta_{\rm crit}$ above which the disc is stable, then we should see the $\beta$ at which fragmentation occurs should converge as the resolution and $\delta t$ are increased. If the disc is also able to fragment stochastically at longer cooling times, then the convergence to a $\beta_{\rm crit}$ can only be found over a limited range of $\delta t$ because if we wait for long enough at a given $\beta$ then fragmentation must eventually occur. However, we can also turn this argument around: if some kind of convergence is obtained over a reasonable range of $\delta t$ then we can place limits on the fragmentation probability at larger $\beta$.

The results of such experiments conducted in boxes of size $16 H$ are given in \autoref{Tab:Frag}, which do indeed suggest convergence of the critical cooling time at resolutions above 8 cells per scale-height. We find that the $\beta$ at which fragmentation occurs converges to $\beta_{\rm crit}\approx4.0$--5.0 for $\delta t \gtrsim 8\upi$, although there is some scatter within that range. We note that this threshold is slightly higher than the $\beta_{\rm crit} \approx 3$ found recently by \citet{Deng2017} and \citet{Baehr2017}. In the following we will interpret this difference as being due to the role of stochastic fragmentation.

To explore the role of stochastic fragmentation, we turn to the second set of experiments. For these experiments we take the simulations presented in \autoref{Tab:SimStats} with $L \le 32H$ and immediately reduce the cooling time $\beta$. We focus on $L \le 32H$ to allow the comparison between three different resolutions. The simulations are then run for another $100\Omega^{-1}$ or until they fragment, depending on which occurs first. The results are shown in \autoref{Tab:FragDirect}.

\begin{figure}
\includegraphics[width=\columnwidth]{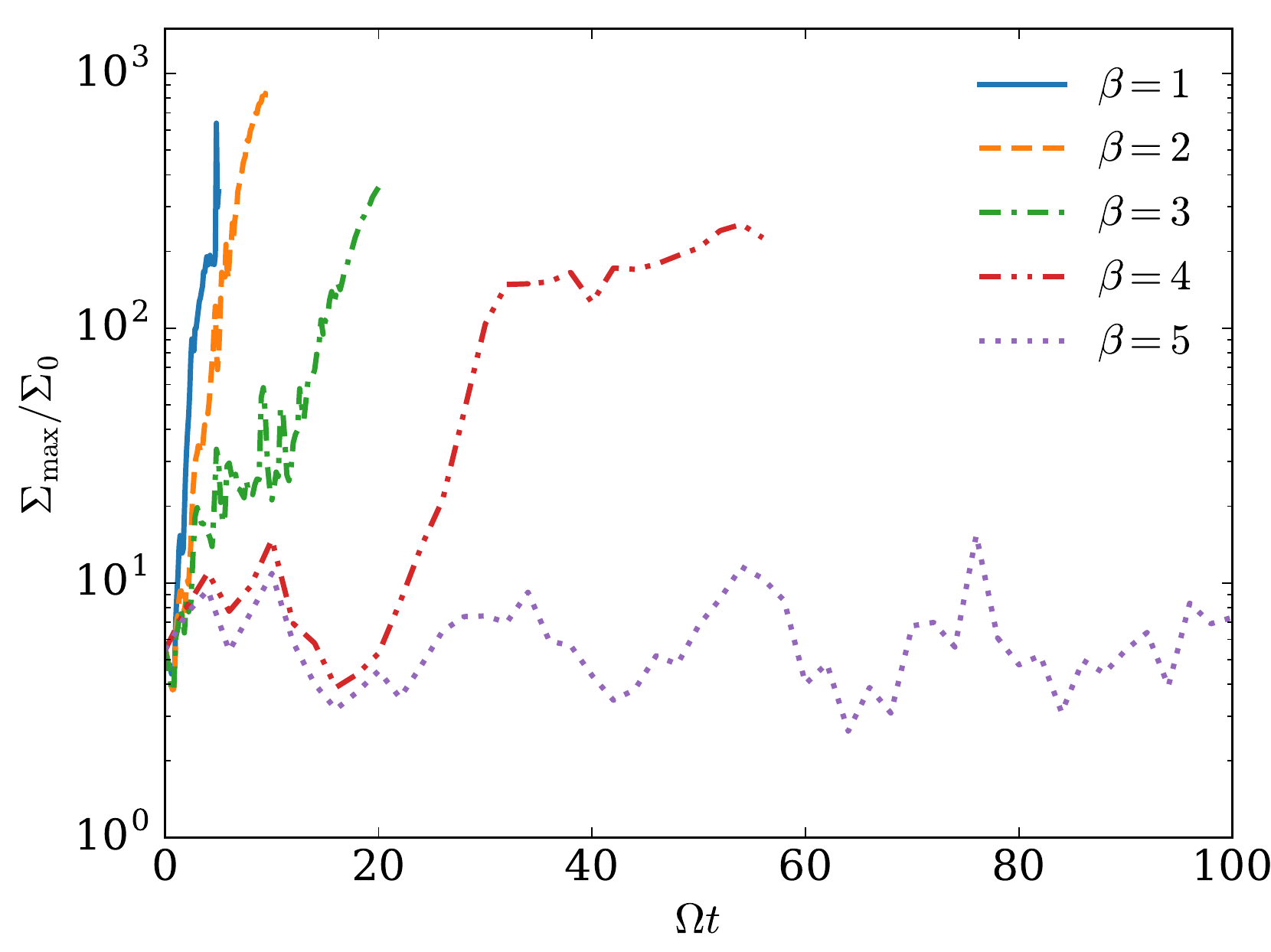}
\caption{Maximum surface density as function of time for different cooling times for the L16N16 simulation. The simulations were initially relaxed into a gravito-turbulent state at $\beta = 10$ before the cooling time was reduced. For $\beta \le 3$ the simulations fragmented on a cooling time, while for $\beta > 3$ they relaxed into a new gravito-turbulent state, from which we observe the formation of a fragment in the $\beta = 4$ simulation.}
\label{Fig:Stochastic}
\end{figure}

Firstly, we note that all simulations with $\beta \le 2$ fragmented immediately. This result is unsurprising given that over densities cool fast enough to allow collapse on a free-fall time for $\beta < \sqrt{2 \upi Q} / (5\gamma-4) \approx 3$, preventing pressure support from stabilising the disc on small scales \citep{Kratter2011}. For $\beta = 3$ we also see that the simulations fragment on a cooling time (excluding the L32N8 run, which is likely affected by low resolution, but still fragments after $50\Omega^{-1}$). However, the borderline nature of $\beta = 3$ is evident in \autoref{Fig:Stochastic}, where a clump with surface density $\Sigma / \Sigma_0 \approx 20$ goes through a phase of relatively steady contraction for around $10\Omega^{-1}$ before rapidly collapsing. 

As in the simulations where the cooling time was slowly reduced, we see fragmentation at $\beta = 4$ given sufficient resolution. However, the mode of fragmentation differs from that seen for $\beta \lesssim 3$, with the disc settling down into a new quasi-steady equilibrium on a shorter time-scale than fragmentation. We note that the length of time the L16N16 simulation took to fragment, 10 to $20\Omega^{-1}$, is similar to the results obtained by slowly reducing the cooling time, which fragmented near $\beta = 4$ when the cooling time was changed over $8 \upi \Omega^{-1} \approx 25\Omega^{-1}$, but was delayed until $\beta \approx 2$ for faster changes.

Although the behaviour seen in \autoref{Fig:Stochastic} hints at the probabilistic, or stochastic, nature of fragmentation, we do not see fragmentation when $\beta > 5$. Our ability to place strict limits on the importance of stochastic fragmentation is limited by the by the relatively short run time of the simulations considered here. However, it is clear that the probability of fragmentation drops rapidly from $\beta = 3$ to $\beta = 5$. Given that a resolution of 16 cells per scale height appears sufficient to resolve fragmentation at $\beta = 4$, it seems unlikely that the reason we do not see fragmentation at $\beta = 5$ in our highest resolution  simulations (32 cells per scale height) is due to a lack of resolution. 

We note that these simulations may over estimate the importance of stochastic fragmentation, because the box sizes used are smaller than the sizes required for convergence. \autoref{Tab:FragDirect} shows that small boxes fragment more easily (e.g. $\beta = 4$ and 32$H^{-1}$ resolution), which should perhaps be expected from the lower average $Q$ and the large amplitude density fluctuations. Both of these factors make it easier for bound clumps to form, and thus we expect that stochastic fragmentation will be even less important at long cooling time than found here.

In addition, we note that none of our simulations fragmented at $\beta = 10$, with our longest run times around $500\Omega^{-1}$ at resolutions of 16 and 32 cells per scale height\footnote{Our long term run, L64N8\_lt, is likely too low resolution to provide useful constraints}. This is already a significant fraction of the lifetime of the self-gravitating phase ($\sim 0.1\unit{Myr} \approx 2000\Omega^{-1}$ at 50\unit{au}). Thus we argue that stochastic fragmentation is unlikely to significantly shift the fragmentation threshold in protoplanetary discs.

\section{Implication for dust dynamics}

The simulations presented here suggest that self-gravitating discs are characterized by turbulence on small scales. On large scales the vertical motions are unimportant, consistent with two-dimensional turbulence, while on small scales the motions become approximately isotropic. The scale at which the dynamics transitions from two- to three-dimensional behaviour is found to be dependent on height above the mid-plane \citep[see also][]{Riols2017}: at $z = H$ the flow power-spectra are already isotropic on length-scales, $\lambda \approx 2H$, yet at the mid-plane this is almost an order of magnitude smaller, around $0.3H$. 

The small-scale turbulence in self-gravitating discs should be efficient at driving large velocity dispersions in the dust particles, leading to high collision velocities. We estimate the magnitude of the collision velocities, $\Delta v$, using the closed form expressions from \citet{Ormel2007}, 
\begin{equation}
\Delta v^2 = \{2,3\} \langle v_g^2 \rangle t_{\rm s} / t_{\rm L}
\end{equation}
where $\langle v_g^2 \rangle$ is the r.m.s. velocity of the turbulence, $t_{\rm s}$ is the particle stopping time, and $t_{\rm L}$ is the correlation time of the largest eddies. This expression is valid for particles with stopping times in the inertial range of the turbulence (the `intermediate range' of \citealt{Ormel2007}) and the factor 2 corresponds to particle pairs with equal $t_s$ while the factor 3 is for the case when $t_{\rm s} \rightarrow 0$ for the smaller particle. Using the relations between $t_{\rm L}$, $E(k)$ and $\langle v_g^2 \rangle$ from \citet{Ormel2007}, the factor 
\begin{equation}
\langle v_g^2 \rangle / t_{\rm L} = \sqrt{18 [k^{5/3}E(k)]^3 }.
\end{equation}
Taking $ k^{5/3} E(k) \approx 0.1$ (\autoref{Fig:Kolmogorov}), we find that the relative velocity in the limit where one particle has negligible $t_s$ is given by $\Delta v \approx 0.4 \sqrt{St}$, where $St = t_{\rm s} \Omega$ is the Stokes number of the larger particle. 

For particles with very different stopping times, the above estimate of the turbulent velocities is in good agreement with the numerical results of \citet{Shi2016} (their Figure~8). For small particles, \citet{Shi2016} showed that the gravitational contribution to the velocity dispersion is small, with the drag coupling to the gas velocity dominating the driving. While the velocity dispersion of  \citet{Shi2016} is roughly a factor 2 smaller than the estimate from the \citet{Ormel2007} estimates derived from power-spectrum, the \citet{Volk1980} models, on which the expressions of \citet{Ormel2007} are based, are known to over-estimate the turbulent velocities by a similar amount \citep{Pan2015}. Thus the collision velocities of small particles in gravito-turbulence are consistent with driving by small-scale turbulence. 

Conversely, for like particles with Stokes numbers between 0.1 and 1, the collision velocity determined from simulations was found to be as much as an order of magnitude smaller than estimate from the power-spectra \citep{Booth2016,Shi2016}. This suggests that the motions of these particles are correlated, which can also be inferred from the narrow filamentary structures for particles of those sizes. The concentration of particles to small regions provides a simple explanation as to why the collisions velocity should be so low: velocity fluctuations on scales larger than the particles' separation should drive correlated motions but not relative motion. As a simple estimate, if one uses the \citet{Ormel2007} formulae,  only including contributions from wavelengths smaller than the width of the filament, about $0.2H$ for $St=1$ \citep{Booth2016}, we find collision velocities that are already a factor $\sim 10$ smaller, closer to the values found from 2D simulations. However, since $0.2H$ is already close to the dissipation scale of the highest resolution simulations presented here, it is likely that in previous simulations the velocity dispersions of particles with $St = 0.1$ and 1 were under resolved, likely resulting in too narrow filaments and also too low collision velocities. 

While it is clear that the small-scale turbulence plays an important role in determining the dynamics of small dust particles, high resolution simulations including dust are required to assess the impact for the largest particles with $St \sim 0.1$ to 1 due to their correlated motions. This is needed to determine not only relative velocity of the particles but also the densities achieved, to asses the viability for planetesimal formation.

\section{Discussion \& Conclusion}

We present the results of a systematic investigation into the stability and properties of gravito-turbulent protostellar discs using three-dimensional shearing box simulations. We find that the general properties of gravito-turbulence are well behaved, and are not sensitive to resolution. However, there is a box-size dependence of the modes beneath a minimum size, $\sim 60H$, which we interpret as being due to the preferred azimuthal wavelength of the spiral modes being around $\sim 60H$.  We attribute the bursty behaviour manifest in our smaller boxes as being due to a large number of long-wavelength modes, which all play a role in the gravito-turbulent dynamics, being missing. We furthermore argue that in global simulations the finite disc size plays a qualitatively similar role in removing such modes when $60 H \sim 2 \pi R$ and note that this is in good quantitative agreement with the results of \citet{Lodato2005} who found an onset of bursty behaviour for disc to star mass ratio above $\sim 0.25$.

At long cooling times we recovered the small-scale turbulence found by \citet{Riols2017} and demonstrated that it has a power-spectrum close to the classical $k^{-5/3}$ result for weakly compressible turbulence. On large scales these motions are largely two-dimensional, becoming more isotropic to smaller scales, although even at the highest resolution presented (64 cells per scale height) the power in azimuthal modes remains lower than the vertical or radial components. We suggest that this turbulence may play an important role in the dynamics of particles in self-gravitating discs, but that turbulence on scales around $0.1H$ will need to be resolved to capture the dynamics.

We explored the stability of gravito-turbulence against fragmentation, supporting previous results that there is a transition between immediate fragmentation and quasi-stable gravito-turbulence at cooling times of around three dynamical times \citep{Gammie2001,Deng2017,Baehr2017}. However, we also find that stochastic fragmentation may be possible at cooling times slightly longer than this, up to cooling times around 5 dynamical times. At longer cooling times, we do not see any evidence for fragmentation, if the disc is allowed to relax into quasi-steady gravito-turbulence before hand. We note that due to the larger amplitude density fluctuations for smaller box sizes, it is likely that our results represent an upper limit to the practical range of cooling times in which fragmentation can occur, rather than the opposite, and thus fragmentation will be limited to the outer regions of protostellar discs.

\section*{Acknowledgements}

We thank Henrik Latter, Antoine Riols and Phil Armitage for numerous discussions throughout this project. This work has been supported by the DISCSIM project, grant agreement 341137 funded by the European Research Council under ERC-2013-ADG and has used the DIRAC Shared Memory Processing and DiRAC Data Analytic systems at the University of Cambridge. The DIRAC Shared Memory Processing system is operated by the COSMOS Project at the Department of Applied Mathematics and Theoretical Physics and was funded by BIS National E-infrastructure capital grant ST/J005673/1, STFC capital grant ST/H008586/1. The DiRAC Data Analytic system was funded by BIS National E-infrastructure capital grant ST/J005673/1, STFC capital grant ST/H008586/1. Both systems are on behalf of the STFC DiRAC HPC Facility (www.dirac.ac.uk), funded by the STFC DiRAC Operations grant ST/K00333X/1. The analysis done in this project made use of the SciPy stack \citep{Scipy}, including NumPy \citep{Numpy}  and Matplotlib \citep{Matplotlib}.

\bibliography{gravoturb}
\bibliographystyle{mnras_edit}

\bsp

\label{lastpage}
\end{document}